\documentclass[12pt,preprint ]{aastex}

\begin{document}

\title{Neutrino Oscillations Induced by Chiral Phase Transition
       in a Compact Star}

\author{\normalsize{Chengfu Mu,
Gaofeng Sun and Pengfei Zhuang}}

\affil{Physics Department,Tsinghua University, Beijing 100084,
China}

\begin{abstract}

 Electric charge neutrality in a compact star provides an
important relationship between the chiral dynamics and neutrino
propagation in the star. Since the sudden drop of the electron
density at the critical point of the first-order chiral phase
transition, the oscillation for low energy neutrinos is significant
and can be regarded as a signature of chiral symmetry restoration in
the core of the star.

\end{abstract}


\section {Introduction}
\label{s1}
It is generally believed that there are two important QCD phase
transitions in hot and dense nuclear matter. One of them is related
to the deconfinement process in moving from a hadron gas to a
quark-gluon plasma, and the other one describes the transition from
the chiral symmetry breaking phase to the phase in which it is
restored. Furthermore, it appears from lattice simulations of QCD
that both transitions coincide at a temperature of about $T=165$ MeV
at zero baryon density(Hwa 1990). The transitions may happen in the
early universe and the early stage of relativistic heavy-ion
collisions where the temperature is expected to be extremely high.
The other often considered systems to realize the transition
condition are compact stars where the temperature is low but the
baryon density can reach several times the normal nuclear density
(Glendenning 2000). From the recent lattice QCD simulations, the
phase structure at high density is much more rich than that at high
temperature. Apart from the quark deconfinement and chiral
restoration, there may be pion superfluidity (Kogut \& Sinclair
2002;He,Jin,\& Zhuang 2005) and even color superconductivity
(Alford,Rajagopal,\& Wilczek 1998)in compact stars. Different from
the temperature effect which leads to a second order phase
transition or even only a crossover, the QCD phase transitions at
high baryon density are all of first order (Karsch 2002) which can
help us to extract signatures in experiments easily.

The signatures of a QCD phase transition that happened in a compact
star depends strongly on the structure of the star, namely the
equation of state of the star. Normally, a compact star is
considered to be completely in a new state of matter, and a bag
constant is needed to balance the pressure on the surface of the
star (Glendenning 2000). In this case, there is no phase transition
inside the star, and the signatures reflect only the properties of
the new state of matter. If a phase transition occurs inside a
compact star, the sudden change in the equation of state at the
critical point of the first order phase transition may lead to some
easily observable signatures. These signatures are connected to the
phase transition itself and different from those related only to the
new state of matter.

Neutrinos play an important role in the study on compact stars
(Bethe 1990;,Qian et al.1993;Prakash et al.1997;Yakovlev et
al.2001). In this paper, we investigate the neutrino conversion in a
compact star in the frame of an effective QCD model with chiral
symmetry. While it is generally believed that the neutrino
oscillation is difficult to occur in compact stars (Bethe 1990), the
situation may be changed when the transition from chiral restoration
phase to chiral breaking phase is taken into account. We will show
that a sudden drop of the electron density associated with the first
order chiral phase transition leads to an remarkable neutrino
oscillation.

The paper is organized as follows. In Section \ref{s2} we describe
the chiral thermodynamics in the Nambu--Jona-Lasinio (NJL) model in
mean field approximation and then by considering the electric charge
neutrality constraint we associate the chiral properties with the
inner structure of a compact star via the Tolman-Oppenheimer-Volkoff
(TOV) equation. In Section \ref{s3} we study the effect of chiral
phase transition on the neutrino oscillation through the neutrino
propagation in the star. We summarize and discuss the obtained
results in Section \ref{s4}. We use the natural unit of
$c=\hbar=k_B=1$ through the paper.

\section {Chiral Thermodynamics}
\label{s2}
The study on QCD phase diagrams depends on lattice QCD calculations
and effective models with QCD symmetries. Since there is not yet
precise lattice result at finite baryon density due to the fermion
sign problem(Karsch 2002), the structure of chiral phase transition
at moderate baryon density is mainly investigated in many low energy
effective models. One of the models that enables us to see directly
how the dynamical mechanisms of chiral symmetry breaking and
restoration operate is the NJL model(Nambu \& Jona-Lasinio 1961)
applied to quarks (Vogl \& Weise 1991;Klevansky 1992;Volkov
1993;Hatsuda \& Kunihiro 1994; Buballa 2005). Recently, the trapped
neutrino effect on the structure of neutron star was investigated in
this model(Menezes,Providencia,\& Melrose 2006; Ruster et al. 2006).
We describe the equation of state of the start with the two flavor
NJL model defined by the Lagrangian density(Vogl \& Weise
1991;Klevansky 1992;Volkov 1993;Hatsuda \& Kunihiro 1994;Buballa
2005)
\begin{equation}
\label{njl} {\cal L}=\bar{q}\left(i\gamma^{\mu}\partial_{\mu}-
m_{0}+\bar\mu\gamma_0\right)q+G\left[\left(\bar{q}q\right)^2+\left(\bar{q}i
\gamma_{5}\vec{\tau}q\right)^2\right],
\end{equation}
where $m_0$ is the current quark mass, $G$ the effective coupling
constant, and $\bar\mu$ the chemical potential matrix in flavor
space,
$\bar\mu=diag\left(\mu_u,\mu_d\right)=diag\left(\mu_B/3-2\mu_e/3,\mu_B/3+\mu_e/3\right)$
with $\mu_B$ and $\mu_e$ being the baryon and electron chemical
potential, respectively.

The essential quantity characterizing a system in a grand canonical
ensemble can be taken to be the thermodynamical potential $\Omega$.
In mean field approximation, it can be expressed in terms of the
effective quarks(Zhuang,Hufner,\& Klevansky 1994),
\begin{eqnarray}
\label{omegaq} \Omega_q &=& {(m_q-m_0)^2\over 4G}-6\int{d^3{\bf
p}\over
(2\pi)^3}\bigg[2E_q\\
&&+T\ln\left[\left(1+e^{-(E_q+\mu_u)/T}\right)\left(1+e^{-(E_q-\mu_u)/T}\right)\right]\nonumber\\
&&+T\ln\left[\left(1+e^{-(E_q+\mu_d)/T}\right)\left(1+e^{-(E_q-\mu_d)/T}\right)\right]\bigg],\nonumber
\end{eqnarray}
where $m_q=m_0-2G\left<\bar q q\right>$ is the constituent quark
mass, $\left<\bar q q\right>$ the order parameter of the chiral
phase transition, and $E_q=\sqrt{m_q^2+{\bf p}^2}$ the effective
quark energy. To ensure the electric charge neutrality, we should
include electrons in the system. Taking into account their
contribution to the thermodynamic potential,
\begin{eqnarray}
\label{omegae} \Omega_e &=& - 2T \int {d^3{\bf p}\over
(2\pi)^3}\ln\Big[\left(1+e^{-(E_e-\mu_e)/T}\right)\nonumber\\
&&\times\left(1+ e^{-(E_e+\mu_e)/T}\right)\Big]
\end{eqnarray}
with the electron energy $E_e=\sqrt{m_e^2+{\bf p}^2}$ and electron
mass $m_e$, the total thermodynamic potential of the system is
\begin{equation}
\label{omegat} \Omega=\Omega_q + \Omega_e.
\end{equation}

The physical order parameter $\left<\bar q q\right>$ or the
effective quark mass $m_q$ should correspond to the minimum of the
thermodynamic potential,
\begin{equation}
\label{gap} {\partial\over\partial
m}\Omega\left(T,\mu_B,\mu_e,m_q\right)=0.
\end{equation}

With the known thermodynamical potential, the pressure $P$ and the
energy density $\epsilon$ of the system are related to $\Omega$ by
\begin{eqnarray}
\label{pe}
P &=&  -\Omega,\nonumber\\
\epsilon &=& -P-T{\partial\Omega\over\partial
T}-\mu_u{\partial\Omega\over\partial\mu_u}-\mu_d{\partial\Omega\over\partial\mu_d},
\end{eqnarray}
where we have considered the electric charge neutrality condition,
namely that we choose the electron chemical potential $\mu_e$ such
that the system has zero net electric charge(Huang,Zhuang,\& Chao
2003),
\begin{equation}
\label{neutrality} n_e =
-{\partial\over\partial\mu_e}\Omega\left(T,\mu_B,\mu_e,m_q\right)=0.
\end{equation}

The coupled set of gap equation and neutrality equation determines
self-consistently the quark mass $m_q$ and the electron chemical
potential $\mu_e$ as functions of temperature $T$ and baryon
chemical potential $\mu_B$. Since we focus on compact stars, the
temperature effect can be safely neglected.

\begin{figure}[!htb]
\begin{center}
\includegraphics[width=6cm]{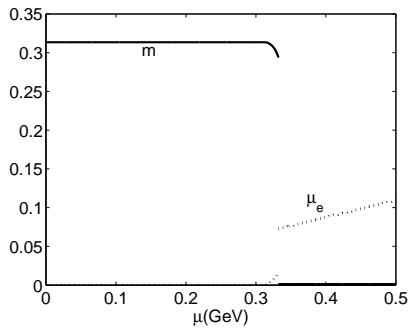}
\caption{The effective quark mass $m_q$ (solid line) and electron
chemical potential $\mu_e$ (dashed line) as functions of
$\mu=\mu_B/3$ at zero temperature in chiral limit of the NJL model.}
\label{fig1}
\end{center}
\end{figure}

To simplify the calculation, we consider in the following chiral
limit of the NJL model with $m_0=0$. There are only two parameters
in the model, the coupling constant $G$ and the hard momentum cutoff
$\Lambda$ to regulate the model. By fitting the pion decay constant
and chiral condensate in the vacuum, they are fixed to be $G=5.02$
GeV$^{-2}$ and $\Lambda=653$ MeV(Zhuang et al.1994). With these
values, the effective quark mass $m_q$ and electron chemical
potential $\mu_e$ are shown in Fig.\ref{fig1} as functions of
$\mu=\mu_B/3$ at zero temperature. Since $m_q$ can be considered as
the order parameter of chiral phase transition, its jump from $294$
MeV to zero at $\mu_c= 332$ MeV means a first order chiral phase
transition. The system is in the chiral symmetry breaking phase at
$\mu<\mu_c$ and the symmetry restoration phase at $\mu
> \mu_c$. The corresponding critical energy density at the transition is $\epsilon_c=0.3$ GeV/fm$^3$.
Due to the electric charge neutrality constraint, $\mu_e$ jumps up
at the transition point and is approximately proportional to $\mu$
in the symmetric phase.

With the above equation of state with chiral phase transition, the
structural of a non-rotating compact star can be obtained by
integrating the TOV equations(Glendenning 2000),
\begin{eqnarray}
\label{tov} &&\frac{d P}{d r}=-G_N
\frac{\left(\epsilon+P\right)\left(M+4\pi r^3P\right)}{r\left(r-2G_NM\right)},\nonumber\\
&&\frac{d M}{d r}=4\pi r^2\epsilon,
\end{eqnarray}
where $G_N=6.707\times10^{-39}$ GeV$^{-2}$ is the universal constant
of gravitation and $M(r)$ the gravitational mass enclosed within the
radius $r$. The chiral properties of the star is reflected in the
pressure $P(r)$ and energy density $\epsilon(r)$. Giving a central
pressure $P$ at $r=0$ with $\mu
> \mu_c$ which ensures that the center of the star is in the chiral restoration phase,
one obtains the $r$-dependence of $P$, and the radius $R$ of the
star is determined by the condition $P(R)=0$. If the star is assumed
to be completely in the chiral restoration phase, one needs a bag
constant to balance the nonzero pressure on the surface of the star.
Taking the bag constant $B=75$ MeV/fm$^3$, the maximum radius of the
star is $R=7.1$ km corresponding to the initial energy density
$\epsilon(0)=1.5$ GeV/fm$^3$ and final energy density
$\epsilon(R)=0.6$ GeV/fm$^3>\epsilon_c$. In our case with chiral
phase transition inside the star, the core of the star is in the
chiral restoration state and it is surrounded by the chiral breaking
state. Taking into account the fact from the lattice simulations
that the deconfinement and chiral restoration happen at the same
critical point, the core can be regarded as a quark matter and the
surrounding a hadron matter. For the initial energy density
$\epsilon(0)=0.32$ GeV/fm$^3$, the chiral phase transition occurs at
the critical radius $R_c=5.2$ km.

\section {Neutrino Oscillations}
\label{s3}
The neutrino oscillation in matter is very different from the
oscillation in vacuum (Caldwell,Fuller,\& Qian 2000; Wolfenstein
1978;Mikheyev \& Smirnov 1985;Rosen \& Gelb 1986;Bilenky \& Petcov
1987; Gelb,Kwong,\& Rosen 1997; Barget, Phillips, \& Whisnant 1986;
Kuo \& Pantaleone 1989; Qian \& Fuller 1995; Savage,Malaney,\&
Fuller 1991; Notzold \& Raffelt 1988; Botella et al.1987). We
discuss in this paper only the $\nu_e\rightarrow \nu_\tau$
oscillation in a compact star with chiral phase transition. In this
case, the neutrino propagation in the star can be expressed as (Kuo
\& Pantaleone 1989)
\begin{eqnarray}
\label{oscillation}
&&i\frac{d}{dr}\left(\begin{array}{c}\nu_{e}\\\nu_{\tau}\end{array}\right)=
\frac{1}{4E}\bigg[(\Sigma+A)+\\
&&\ \ \ \ \ \ \ \ \ \ \ \ \ \ \ \left(\begin{array}{cc}A-\Delta\cos
2\theta&\Delta\sin 2\theta\\\Delta\sin 2\theta&-A+\Delta\cos
2\theta\end{array}\right)\bigg]\left(\begin{array}{c}\nu_{e}\\\nu_{\tau}\end{array}\right)\nonumber
\end{eqnarray}
where $\Sigma=m_2^2+m_1^2$ and $\Delta=m_2^2-m_1^2$ are related to
the neutrino mass eigenvalues $m_1$ and $m_2$ in vacuum, and
$\theta$ is the mixing angle between the neutrino mass eigenvalue
and the weak interaction eigenvalue. Since $\Sigma$ contributes only
a global phase factor to the neutrino eigenstate and does not affect
the neutrino conversion(Fukugita \& Yanagida 2003), we omit it in
the following. The matter dependence of the conversion is reflected
in the factor $A$ defined as
\begin{equation}
\label{a} A=2\sqrt 2 G_F E N_e,
\end{equation}
where $E$ is the neutrino energy, $G_F=1.166\times 10^{-5}$
GeV$^{-2}$ the Fermi constant, and $N_e$ the free electron density
\begin{equation}
\label{ne} N_e(r) = 2\int{d^3{\bf p}\over (2\pi)^3}{1\over
1+e^{(E_e-\mu_e(r))/T}}.
\end{equation}
the $r-$dependence of the electron chemical potential is given by
the TOV equation together with the equation of state. Suppose only
electron neutrinos are created at the center of the star, the
combination of the propagation equation (\ref{oscillation}), the TOV
equation (\ref{tov}) and the chiral equation of state (\ref{pe})
controls self-consistently the space evolution of the transition
from electron to tau neutrinos.

To find the most suitable electron density $N_e$ where there is
remarkable neutrino oscillation, we first consider the case with
space-independent $\mu_e$. In this case, the propagation equation
(\ref{oscillation}) can be analytically solved, and we obtain the
explicit surviving probability
\begin{equation}
P_e(r)=1-\frac{1}{2}\sin^2 2\theta_m\left(1-\cos{2\pi r\over
\lambda_m}\right),
\end{equation}
where $\sin^2 2\theta_m$ and $\lambda_m$ depend on $\mu_e$ via
\begin{eqnarray}
\label{sin} \sin^2 2\theta_m &=& {(\Delta\sin 2\theta)^2\over
(-A+\Delta\cos2\theta)^2+(\Delta\sin2\theta)^2},\nonumber\\
\lambda_m &=& {4\pi E\over \sqrt{(\Delta\sin 2\theta)^2 +(A-\Delta
\cos 2\theta)^2}}.
\end{eqnarray}
In vacuum with $N_e=0$, the surviving probability is reduced to
\begin{equation}
P_e(r)=1-\frac{1}{2}\sin^2 2\theta\left(1-\cos\frac{\Delta
r}{2E}\right).
\end{equation}

It is easy to see that the biggest transition in matter happens at
$\sin^22\theta_m =1$ which corresponds to the matter environment
with resonant electron density
\begin{equation}
\label{ne0} N_{er}={\Delta\cos 2\theta\over 2\sqrt 2 G_F E},
\end{equation}
where $\Delta, \theta$ and $G_F$ are parameters fixed in the vacuum,
the only adjustable parameter is the neutrino energy $E$. For low
energy neutrinos, the resonant electron density $N_{er}$, or
equivalently, the resonant electron chemical potential $\mu_{er}$ is
large, and the matter induced oscillation is remarkably different
from the oscillation in vacuum. With increasing neutrino energy,
$N_{er}$ approaches to zero, and the difference in the oscillation
between in matter and in vacuum disappears gradually.

While the resonant electron chemical potential $\mu_{er}$ is
independent of the chiral thermodynamics and the star structure, the
matter effect to the neutrino oscillation is sensitive to the
initial condition of the TOV equation, namely the equation of state
at $r=0$. For a smooth $r-$dependence of $\mu_e(r)$ around
$\mu_{er}$ characterized by the TOV equation, the $r-$integrated
matter contribution to the neutrino oscillation is visible, while a
sharp $\mu_e(r)$ around $\mu_{er}$ will make the matter effect to
the neutrino oscillation difficult to be seen.

\begin{figure}[!htb]
\begin{center}
\includegraphics[width=6cm]{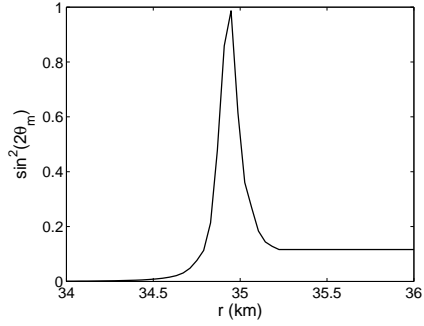}
\caption{The effective neutrino mixing angle $\sin^2 2\theta_m$ as a
function of radius $r$ of the star with initial energy density
$\epsilon(0)=0.32 $ GeV/fm$^3$ for low energy neutrinos with energy
$E=0.1$ MeV.}
 \label{fig2}
\end{center}
\end{figure}

Taking the vacuum parameters $\Delta=2.6\times 10^{-3}$ eV$^2$ and
$\sin^2\theta=0.03$(Maltoni 2004;Apollonio 1999), the effective
neutrino mixing angle in matter is shown in Fig.\ref{fig2} as a
function of radius $r$ for low energy neutrinos with $E=0.1$ MeV.
For the initial energy density $\epsilon(0)=0.32$ GeV/fm$^3$, the
peak $\sin^2 2\theta_m =1$ is located at the resonant radius
$R_r\sim 35$ km $>R_c$ in the chiral symmetry breaking phase. This
means that a remarkable matter induced neutrino oscillation can not
happen in a compact star without chiral phase transition, since the
electron density $N_e$ in chiral restoration phase is too large and
far away from the resonant electron density $N_{er}$.

\begin{figure}[!htb]
\begin{center}
\includegraphics[width=6cm]{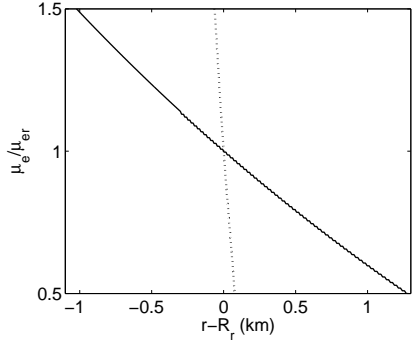}
\caption{The electron chemical potential $\mu_e$ scaled by its
resonant value $\mu_{er}$ as a function of $r-R_r$ with $R_r$ being
the resonant radius where the biggest neutrino transition occurs.
The solid and dashed lines correspond, respectively, to the initial
energy density $\epsilon(0)=0.32$ and $0.39$ GeV/fm$^3$. }
 \label{fig3}
\end{center}
\end{figure}

To see the initial energy density dependence of the neutrino
oscillations, we plot the electron chemical potential $\mu_e$ around
$\mu_{er}$ as a function of $r$ in Fig.\ref{fig3} for different
initial energy densities. The slope is very sensitive to the value
of $\epsilon(0)$. For $\epsilon(0)=0.32$ GeV/fm$^3$, $\mu_e$ varies
smoothly in the neighborhood of $\mu_{er}$, and therefore, the space
integrated matter contribution to the neutrino oscillation can be
seen clearly. For $\epsilon(0)=0.39$ GeV/fm$^3$, $\mu_e$ changes
very fast around $\mu_{er}$, and the matter contribution occurs only
in a very thin spherical shell of the star. In this case, the
$r-$integrated matter effect is negligible.

\begin{figure}[!htb]
\begin{center}
\includegraphics[width=6cm]{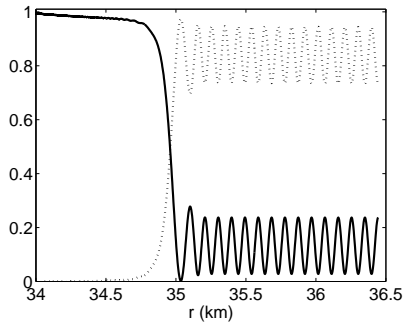}
\caption{The surviving probability $P_e$ of electron neutrinos
(solid line) and their conversion probability $P_\tau$ to $\tau$
neutrinos (dashed line) as functions of $r$ for initial energy
density $\epsilon(0)=0.32$ GeV/fm$^3$ and neutrino energy $E=0.1$
MeV. }
 \label{fig4}
\end{center}
\end{figure}

In Fig.\ref{fig4} we show the surviving probability (solid lines) of
the electron neutrinos and their transition probability to tau
neutrinos (dashed line) as functions of $r$. From our assumption of
no tau neutrinos at the center of the star, the surviving
probability $P_e(r)$ starts propagation with the initial value
$P_e(0)=1$. In the core with chiral symmetry restoration, there is
almost no neutrino conversion due to the too dense electrons. At the
critical radius $R_c$ where the phase transition from chiral
restoration to chiral breaking happens, the electron chemical
potential and in turn the electron density drops down suddenly to a
small value, and the visible conversion starts gradually. The
biggest change occurs around the resonant radius $R_r$. For low
energy neutrinos with $E=0.1$ MeV, almost all the electron neutrinos
are converted into tau neutrinos around $R_r$, and this conversion
is kept in the further propagation. Clearly, the conversion
probability is $P_\tau(r)=1-P_e(r)$. With increasing neutrino
energy, the conversion is suppressed by the too large resonant
radius in the chiral breaking state, as we explained above. For
$E\sim 0.3$ MeV, the maximum change from electron to tau neutrino is
reduced to about $50\%$, and for $E>0.8$ MeV, there is almost no
difference between the neutrino oscillations in matter and in
vacuum.

\section {Summary and Discussion}
\label{s4}
We have examined the $ \nu_e\rightarrow\nu_\tau $ neutrino
oscillation induced by chiral phase transition in a compact star.
Due to the constraint of electric charge neutrality, the chiral
properties controlled by baryon chemical potential $\mu_B$ and the
neutrino conversion governed by electron chemical potential $\mu_e$
are related to each other. The star contains a dense core with
chiral symmetry restoration rounded by the chiral breaking phase.
When neutrinos propagate in the radius direction, the transition
from chiral restoration to chiral breaking inside the star leads to
significantly oscillations for low energy neutrinos. When the
neutrino energy is low enough, the conversion probability is nearly
$100\%$. Therefore, it is possible to detect the inner structure of
compact stars in terms of low energy neutrino oscillations.

We emphasize that the remarkable oscillation is resulted from the
first-order chiral phase transition. For a second-order phase
transition, due to the lack of the sudden drop of the electron
chemical potential, the resonant radius $R_r$ will be several
hundreds or even several thousands kilometers which are certainly
beyond the reasonable radius of a compact star. In our calculation,
$R_r$ is about $35$ kilometers for initial energy density
$\epsilon(0)=0.32$ GeV/fm$^3$ and is reduced to about $13$
kilometers for $\epsilon(0)=0.39$ GeV/fm$^3$. Since the first-order
chiral phase transition is a general consequence of lattice
simulations and model calculations at high baryon density, our
results obtained in the frame of NJL model is still qualitatively
correct in general case.

When other phase transitions at high baryon density are taken into
account, what is their effect on the neutrino oscillations? Since
the other transitions such as color superconductivity occur at
extremely high density where the chiral symmetry is already
restored, the electron chemical potential $\mu_e$ in these new
phases is much higher than the resonant value $\mu_{er}$, and
therefore, these new phases will not lead to remarkable changes in
the neutrino oscillations. Similarly, the nonzero current quark mass
$m_0$ will change the behavior of $\mu_e$ in the chiral restoration
phase considerably, but it changes the neutrino conversion slightly.
The neutrino oscillation in compact star is characterized by the
first-order chiral phase transition.

{\bf Acknowledgments:}\  The work was supported in part by the
grants NSFC10575058, 10425810, 10435080 and SRFDP20040003103.

\end{document}